%% file: main.tex
\documentclass[num-refs]{wiley-article}

\usepackage{siunitx}
\usepackage{float}

\papertype{Original Article}
\paperfield{Journal Section}

\title{Comparing restricted mean survival times in small sample clinical trials using pseudo-observations }

\author[1]{David Jesse}
\author[1]{Cynthia Huber}
\author[1]{Tim Friede}

\affil[1]{Department of Medical Statistics, University Medical Center Göttingen, Göttingen, Germany}

\corraddress{David Jesse, Department of Medical Statistics, University Medical Center Göttingen, 37075 Göttingen, Germany}
\corremail{david.jesse@gmx.net}

\fundinginfo{The authors are grateful for funding by the DFG grant number FR 3070/4-1.}

\runningauthor{Jesse et al.}

\input{operators}
\input{reffuns}

\begin{document}

\begin{frontmatter}
\maketitle

\begin{abstract}
The widely used proportional hazard assumption cannot be assessed reliably in small-scale clinical trials and might often in fact be unjustified, e.g. due to delayed treatment effects. 
An alternative to the hazard ratio as effect measure is the difference in restricted mean survival time (RMST) that does not rely on model assumptions. 
Although an asymptotic test for two-sample comparisons of the RMST exists, it has been shown to suffer from an inflated type I error rate in samples of small or moderate sizes. 
Recently, permutation tests, including the studentized permutation test, have been introduced to address this issue. 
In this paper, we propose two methods based on pseudo-observations (PO) regression models as alternatives for such scenarios and assess their properties in comparison to previously proposed approaches in an extensive simulation study. 
Furthermore, we apply the proposed PO methods to data from a clinical trail and, by doing so, point out some extension that might be very useful for practical applications such as covariate adjustments.

\end{abstract}
\end{frontmatter}

\input{sections/01-intro}
\input{sections/02-methods}
\input{sections/03-simulation}
\input{sections/04-example}
\input{sections/05-conclusion}

\printendnotes

\bibliography{references,packages}

\end{document}

%% file: operators.tex
\newcommand{\xvec}{\mathbf{x}}
\newcommand{\xmat}{\mathbf{X}}
\newcommand{\betavec}{\boldsymbol{\beta}}

\newcommand{\omegamat}{\boldsymbol{\Omega}}
\newcommand{\tvec}{\mathbf{t}}
\newcommand{\deltavec}{\boldsymbol{\delta}}
\newcommand{\fmat}{\mathbf{F}}

\newcommand{\mmat}{\mathbf{M}}

\newcommand{\residhat}{\hat{\varepsilon}}

\newcommand{\betahat}{\hat{\beta}}
\newcommand{\muhat}{\hat{\mu}}

\newcommand{\sehat}{\widehat{\operatorname{se}}}

\newcommand{\Covhat}{\widehat{\operatorname{Cov}}}

\newcommand{\expect}{\operatorname{\mathbb{E}}}
\newcommand{\diag}{\operatorname{diag}}

\newcommand{\real}{\mathbb{R}}

%% file: reffuns.tex
\newcommand{\secref}[1]{Section~\ref{#1}}
\newcommand{\figref}[1]{Figure~\ref{#1}}
\newcommand{\tabref}[1]{Table~\ref{#1}}

%% file: sections/01-intro.tex
\section{Introduction} \label{sec-intro}

In clinical trials with time-to-event endpoints, the log-rank test and the hazard ratio are commonly used for testing and characterising treatment effects, respectively. 
However, they both rely on the assumption that the hazard functions of the treatment groups are proportional over time. 
In case of non-proportional hazards (NPH), the log-rank test loses power and the interpretation of the hazard ratio as the effect measure becomes ambiguous. 
Since violations of the proportional hazards assumption have been increasingly observed in practice, for instance in oncology trials evaluating immunotherapies with delayed treatment effects, a range of methods was developed taking such violations into account \cite{bardo2024}. 
One strand of research aims to replace the hazard ratio with model-free effect measures and to conduct statistical inference based on these \cite{quartagno2023}. 
A prominent example of such effect measures is the difference in restricted mean survival time (RMST) \cite{royston2011}. 
While an asymptotic test for two-sample comparisons of the RMST exists and is easy to implement \cite{hasegawa2020}, it has been shown that it suffers from an inflated type I error rate when the sample size is small or moderate \cite{horiguchi2020a}. 
For this reason, adaptations to the RMST-based test have been made to improve its performance for small sample sizes, e.g. by employing permutation methods \cite{horiguchi2020a,ditzhaus2023}. 
In particular, \citet{ditzhaus2023} introduced a studentized permutation test for two-sample RMST comparisons and demonstrated its superiority to the asymptotic test and other suggested alternatives from the literature in terms of type I error rate control. 

Alternatively, RMST-based inference can be based on pseudo-observations (PO) \cite{andersen2010}. 
Motivated by the goal of evaluating covariate effects on arbitrary survival quantities, the idea is to circumvent the need to deal with censored observations and estimate a generalized linear model (GLM). 
For this, pseudo-observations of the given survival quantity are calculated for each individual and are then used as the response variable for fitting a GLM. 
Typically, robust sandwich-type estimators of the covariance matrix of the regression coefficients \cite{zeileis2006} are employed for inference, most commonly the HC3 estimator \cite{sachs2022,mackinnon1985}. 
Since in other contexts, e.g. for linear models with heteroscedastic errors, using this estimator has been shown to exhibit favorable small sample properties \cite{long2000}, we hypothesize that this is also the case for PO regression models for the RMST. 
Although PO methods have gained some popularity in survival analysis, in particular for RMST-based analyses \cite{royston2011,andersen2017,ambrogi2022}, to the best of our knowledge, their usage has not been investigated with regard to their small sample performance. 
In this paper, we want to close this gap and present two PO methods for estimating and testing two-sample RMST differences in \secref{sec-methods}. The first method employs an asymptotic test as it has been discussed in previous research \cite{hasegawa2020}. 
Moreover, we propose a bootstrap method as a potential alternative to the asymptotic test. 
Whilst a bootstrap procedure has been proposed for robust estimation of the covariance matrix \cite{andersen2010,sachs2022}, resampling approaches have not been investigated systematically for hypothesis testing. 
We assess the performance of the proposed approaches in comparison to selected alternatives from the literature \cite{hasegawa2020,ditzhaus2023} in an extensive simulation study in \secref{sec-sim}. 
 In \secref{sec-example} we illustrate the application of these methods using a data set from a clinical trial. 
 The manuscript concludes with remarks on the potential advantages and disadvantages of our proposed methods, see \secref{sec-discuss}.

%% file: sections/02-methods.tex
\section{Methodology} \label{sec-methods}
Let $T \in \real^{+}$ denote a random variable for the event time of interest and $C \in \real^{+}$ a second random variable for the censoring time. 
For an individual $i$, we observe $t_i = \min(T_i, C_i)$ due to right-censoring. 
The censoring indicator $\delta_i = \mathbf{1} \{T_i \leq C_i\}$ records whether the event time $t_i$ has been observed or is censored. 
We consider the two-sample setting in a randomized controlled trial for which we assume survival and censoring times that are identically distributed within each group and mutually independent:
\begin{equation*}
    T_{ji} \sim S_j, \quad C_{ji} \sim G_j \quad (j = 0, 1; \, i = 1, \ldots, n_j)
\end{equation*}
Here, $S_j$ and $G_j$ denote the survival functions of the event and censoring times of group $j$, respectively, and $i \in \{1, \ldots, n_j\}$ indicates the $i$-th individual within that group. 
With $j = 0$, we denote the reference or placebo treatment group, while $j = 1$ indicates the experimental treatment group. 
In the context of PO regression models, it is useful to denote the treatment assignment with an indicator variable $Z \in \{0, 1\}$ rather than using the subscript $j$. Here $Z = 1$ denotes that the experimental treatment has been received. 

The (groupwise) RMST is defined as the integral of the survival function from time $0$ up to the prespecified truncation time point $t^*$:
\begin{equation*}
    \mu_j(t^*) = \expect[\min(T_j,\, t^*)] = \int_0^{t^*} S_j(u) \,du
    \label{eq:test-problem}
\end{equation*}
In terms of its interpretation, this corresponds to the expected survival time of the population when the observation period is restricted to the aforementioned time window. 
An estimate of the RMST can be obtained by plugging in an estimator $\hat{S}_j$ of the survival function: 
\begin{equation*}
    \muhat_j(t^*) = \int_0^{t^*} \hat{S}_j(u) \, du 
\end{equation*}
Any valid estimator of the survival function can be used but typically the Kaplan-Meier estimator \cite{kaplan1958} is employed. 
Due to its non-parametric nature, it protects against making false modeling assumptions. 
Moreover, by employing the Kaplan-Meier estimator we can derive closed-form solutions for the point estimator of the RMST as well as its standard error \cite{hasegawa2020}. 
Given these estimators, we can test whether the RMST difference is different from $0$ for a certain restriction time $t^*$: 
\begin{equation}
    H_0: \mu_1(t^*) - \mu_0(t^*) = 0 \quad \text{vs.} \quad H_1: \mu_1(t^*) - \mu_0(t^*) \neq 0
\end{equation}
A corresponding testing procedure based on asymptotic theory is presented, among others, by \citet{hasegawa2020}. 
Since this method has been shown to suffer from an inflated type I error rate under small and even moderate sample sizes, \citet{ditzhaus2023} developed a studentized permutation test based on previous work by \citet{horiguchi2020a}, which addresses this problem. 
In the next subsection, we present methods based on pseudo-observations for handling this problem. 
Building upon GLM methodology, these approaches differ from the standard asymptotic and the studentized permutation test.

\subsection{Pseudo-observations} \label{sec-methods-po}

PO methods are broadly applicable to research questions concerning time-to-event data and are particularly useful when interest lies in evaluating covariate effects that are not on the hazard scale \cite{andersen2010}. 
In general, the goal is to estimate a GLM of the form
\begin{equation}
    \expect[V_i \,|\, \xvec_i] = g^{-1}(\xvec_i' \betavec)
    \, , 
    \label{eq:po-glm}
\end{equation} 
where $V_i = f(T_i)$ denotes some transformation of the random variable $T_i$, reflecting the estimand we are interested in \cite{sachs2022}. 
For instance, $f(T_i \,|\, t^*) = \min(T_i,\, t^*)$ and the expectation thereof corresponds to the RMST. 
The linear predictor $\xvec_i' \betavec$ contains the covariate effects on that estimand we want to evaluate plus an intercept term. 
In the following, we consider the case where $\xvec_i' \betavec$ equals $\beta_0 + \beta_1 Z_i$, i.e. the linear predictor only contains the treatment effect on the RMST scale. 
The link function $g$ needs to be chosen by the researcher and determines how the covariate effects $\betavec$ on $\expect[V_i | \xvec_i]$ are interpreted. 
In the simplest setting, $g$ is chosen to be the identity link such that the effects $\betavec$ can be interpreted as differences \cite{sachs2022}. 

Since we need to deal with right-censored data, $V_i$ cannot be computed directly for all individuals and the regression model \eqref{eq:po-glm} cannot be estimated. 
This is where pseudo-observations become relevant. 
Instead of using the original time-to-event observations as the response, we can replace them with pseudo-observations that are calculated in the following way \cite{andersen2010,sachs2022}: 
\begin{equation}
    P_i = n \hat{\theta} - (n - 1) \hat{\theta}_{-i}
    \label{eq:po}
\end{equation}
Here, $\hat{\theta}$ denotes a well-behaved marginal estimator of $\theta = \expect[V_i]$. 
This could be any estimator satisfying asymptotic efficiency but, usually, a non-parametric estimator is used. 
Likewise, $\hat{\theta}_{-i}$ is the same kind of estimator but leaving out the $i$-th observation, i.e. using the $i$-th jackknife sample \cite[Chapter 11]{efron1993}. 
The rationale for using these pseudo-observations is to obtain a synthetic sample for which $\expect[P_i] = \expect[V_i]$ holds and for which each observation can be understood as the contribution to the marginal expectation $\expect[V]$ \cite{andersen2003, andersen2010}. 
Therefore, it is also important to note that, for the statistical analysis, the pseudo-observations are calculated and used for all individuals in the data, regardless of whether $V_i$ could be computed directly or not \cite{andersen2010}. 

Estimation of the model \eqref{eq:po-glm} is based on quasi-likelihood methods, also known as generalized estimating equations (GEE) methods \cite[Chapter 5.5]{fahrmeir2013}. 
These estimating equations are given by the quasi-score function 
\begin{equation*}
    s(\betavec) = \sum_{i = 1}^n \xvec_i \frac{h'(\eta_i)}{\sigma^2_i} (P_i - \mu_i)
    \, , 
\end{equation*}
where $\mu_i = g^{-1}(\xvec_i' \betavec)$, $h = g^{-1}$, $\eta_i = \xvec_i' \betavec$ and $\sigma^2_i$ is a working variance structure. 
The key difference to ordinary likelihood approaches is that only the first two moments are specified but no complete distributional assumption is being made. 
Hence, the variance structure is also not determined by the model directly but must be specified by the researcher as a function of the form $\sigma^2(\mu)$. 
The most common and easiest option is to set $\sigma^2_i = \sigma^2$, i.e. choosing a constant variance \cite{sachs2022}. 
The parameter vector $\betavec$ is then estimated by (numerically) finding the root of the generalized estimating equation $s(\hat{\betavec}) = \mathbf{0}$. 

Valid estimation and inference of the model \eqref{eq:po-glm} rely on the assumption of the censoring mechanism being independent of any covariates. 
If this assumption is questionable there are two ways of dealing with this. 
First, if categorical covariates impact the censoring times, the pseudo-observations can be calculated stratified by the different categories \cite{andersen2010}. 
Second, if we also need to consider continuous variables, methods based on inverse probability of censoring weighting (IPCW) can be applied \cite{binder2014,overgaard2019}. 
In this paper, only the stratification method is considered and will be used throughout.

\subsubsection{Asymptotic test} \label{sec-methods-asy}

In the following, we consider the following particular case of the regression model \eqref{eq:po-glm}:
\begin{equation}
    \expect[\min(T_i, t^*) | Z_i] = \mu_i(t^* | Z_i) = \beta_0 + \beta_1 Z_i
    \, . 
    \label{eq:po-glm-rmst}
\end{equation}
Hence, we aim to evaluate the effect of a treatment, $Z$, on the RMST. 
Furthermore, as we specify $g$ to be the identity link function, $\beta_1$ corresponds to the RMST \textit{difference} between the treatment and the control group. 
The testing problem \eqref{eq:test-problem} is then given by replacing $\mu_1(t^*) - \mu_0(t^*)$ with the coefficient $\beta_1$. 
This test can be carried out using the quasi-likelihood equivalent of ordinary likelihood-based inference \cite[Chapter 5.5]{fahrmeir2013}. 
Hence, we can construct the test statistic $Z^{\text{(PO)}} = \hat{\beta}_1 / \sehat(\hat{\beta}_1)$, which asymptotically follows a standard normal distribution under the null hypothesis. 
Therefore, a (two-sided) statistical test can be conducted by comparing $Z^{\text{(PO)}}$ to the $1 - \alpha /2$ quantile of the standard normal distribution: 
\begin{equation}
    \varphi^{(\text{PO})} = \mathbf{1} \{|Z^{(\text{PO})}| > z_{1- \alpha/2} \}
    \label{eq:po-test}
\end{equation}
Similarly, a symmetric $1 - \alpha$ confidence interval is given by:
\begin{equation}
    CI^{(\text{PO})} = \left[ \hat{\beta}_1 \mp z_{1- \alpha/2} \, \sehat(\hat{\beta}_1) \right]
    \label{eq:po-ci}
\end{equation}

An estimate of the standard error $\sehat(\hat{\beta}_1)$ can be obtained from estimating the covariance matrix $\operatorname{Cov}(\hat{\betavec})$. 
Since the working variance $\sigma_i^2$ is likely to be misspecified, sandwich-type estimators of the covariance matrix are used. 
These have the form $\Covhat(\hat{\betavec}) = \hat{\fmat}^{-1} \hat{\mmat} \hat{\fmat}^{-1}$, where we use the short-hand notation $\hat{\fmat} = \fmat(\hat{\betavec})$ for the inverse of the quasi-Fisher information matrix. 
$\hat{\mmat}$, on the other hand, is an empirical version of the "meat" matrix \cite[Chapter 5.5]{fahrmeir2013}. 
A popular class of such sandwich-type estimators are \textit{heteroscedasticity consistent} (HC) covariance matrix estimators \cite{zeileis2006}, for which $\hat{\mmat} = \xmat' \widehat{\omegamat} \xmat$. 
Here, $\xmat$ denotes the design matrix of the regression model and $\widehat{\omegamat} = \diag(w_1, \ldots, w_n)$ is a diagonal matrix of weights depending on the working residuals $\residhat_i$ \cite{zeileis2006}. 
There exist various specifications of $\widehat{\omegamat}$ leading to different HC-type covariance matrix estimators \cite{zeileis2004}. 
In this paper, we focus on the HC3 covariance matrix estimator \cite{mackinnon1985}, for which the weights, $w_i$, are defined as $\frac{\residhat_i^2}{(1 - h_i)^2}$.
In addition to the working residuals, these weights also depend on the leverages $h_i$ of the observations. 
For classical linear models with heteroscedastic errors, the HC3 estimator has been shown to have the best performance among similar alternative covariance matrix estimators in case of small sample sizes \cite{long2000}. 
Moreover, similar results exist for PO regression models, making it the default option in corresponding software packages as well \cite{sachs2022}.

\subsection{Bootstrap test} \label{sec-methods-boot}

Inspired by the studentized permutation test by \citet{ditzhaus2023}, we consider a second approach based on PO regression models using a resampling method. 
Here, we propose a bootstrap procedure instead of a permutation approach as it remains applicable to continuous covariates. 
The procedure in terms of estimating the regression coefficients $\betavec$ and their covariance matrix $\operatorname{Cov}(\hat{\betavec})$ remains the same as described in \secref{sec-methods-po} and \secref{sec-methods-asy}. 
However, we aim to relax the assumption about the test statistic $Z^{\text{(PO)}}$ to be standard normally distributed under the null hypothesis. 
We therefore intend to estimate this distribution directly from the data using a nonparametric bootstrap approach. 

For this, let $(\tvec, \deltavec, \xmat) \equiv \{(t_i, \delta_i, \xvec_i'): \, i = 1, \ldots, n \})$ denote the original random sample. 
Furthermore, let $(\tvec, \deltavec, \xmat)^b \equiv \{(t_i, \delta_i, \xvec_i')^b: \, i = 1, \ldots, n \})$ denote a \textit{bootstrap} sample that has been obtained by sampling with replacement from the original sample. 
For a given data set, we generate such bootstrap samples $B$ times (e.g. $B = 5000$). 
We then estimate the model \eqref{eq:po-glm-rmst} for each of these bootstrap samples using similar methods as described before. 
Finally, we can calculate the bootstrap test statistics $Z^b = |\betahat_1^b - \betahat_1| / \sehat(\betahat_1^b)$ ($b = 1, \ldots, B$) to obtain samples that can be used for estimating the null distribution of $Z^{\text{(PO)}}$ nonparametrically. 
The rationale of the formula for the bootstrap test statistics is that "the estimate of $\beta$ from the bootstrap samples should, on average, be equal to $\betahat$, at least asymptotically" \cite{mackinnon2009} and therefore $Z^b$ should mimic the distribution of $Z^{\text{(PO)}}$ under the null hypothesis. 
Also note that this approach is generally applicable and not restricted to the specific model \eqref{eq:po-glm-rmst} and the coefficient $\beta_1$ therein. 
Using these bootstrap samples, we can now conduct statistical inference in a similar manner as done by \citet{ditzhaus2023} with their studentized permutation test. 
Thus, a two-sided test for a given significance level $\alpha \in (0, 1)$ is given by
\begin{equation}
    \varphi^{b} = \mathbf{1} \{|Z^{(\text{PO})}| > q_{1 - \alpha}^b\}
\end{equation}
with $q_{1 - \alpha}^b$ being the $(1 - \alpha)$-quantile of the bootstrap test statistics. 
Similarly, we can construct a $1 - \alpha$ bootstrap-t confidence interval \cite[Chapter 12.5]{efron1993} based on that quantile:
\begin{equation}
    CI^b = \left[\betahat_1 \mp  q_{1 - \alpha}^b \, \sehat(\betahat_1)\right]
\end{equation} 

There is, however, one obstacle to this approach hindering its practical feasibility: 
Due to the computation of the pseudo-observations, even the asymptotic test \eqref{eq:po-test} incorporates a resampling scheme. 
Applying a bootstrap on top of that implies a nested resampling procedure, which gets computationally more expensive (a) the larger the original sample is and (b) the more bootstrap samples we want to draw. 
Such a problem can become apparent quickly. 
For instance, if we had a sample of size $n = 50$ and wanted to draw $B = 1000$ bootstrap samples, we would need to carry out $n \cdot B = 50,000$ computations in total. 
One way to tackle this problem is to embrace parallel computing capabilities. 
A more subtle approach is to eliminate one of the two resampling levels entirely. 
This is possible by using an approximate version of the pseudo-observations \eqref{eq:po}:
\begin{equation}
\begin{split}
    P_i &= n \hat{\theta} - (n - 1) \hat{\theta}_{-i} \\
    &= \hat{\theta} + (n - 1) (\hat{\theta} - \hat{\theta}_{-i}) \\
    &\approx \hat{\theta} + n \frac{\partial \hat{\theta}}{\partial w_i}
\end{split}
\end{equation}
These approximations are based on the first-order influence function of $\hat{\theta}$ and are known as \textit{infinitesimal jackknife} (IJ) pseudo-observations \cite{parner2023}. 
It is required that the marginal estimator $\hat{\theta}$ can be written as a function of weights $w_i$ attached to each individual $i = 1, \ldots, n$ in the data. 
As pointed out, the advantage over ordinary pseudo-observations obtained by using the jackknife samples is the computational speed. 
The loss in numerical accuracy is negligible even for moderate sample sizes, making IJ pseudo-observations an attractive alternative to ordinary pseudo-observations \cite{parner2023}. 
This is also confirmed in a simulation study reported in supplement S2.

A further potential issue that needs to be discussed is the possibility of drawing a jackknife or a bootstrap sample for which the Kaplan-Meier estimator and therefore the estimator of the RMST is not uniquely defined at the specified restriction time $t^*$. 
This problem appears when, for one sample, the largest event time is smaller than $t^*$ and censored. 
\citet{horiguchi2020a} discussed this issue extensively for their (unstudentized) permutation method and found that a "horizontal extension strategy" is a pragmatic option that works equally well as other suggested alternatives. 
Like \citet{ditzhaus2023} did for their studentized permutation test, we adopt this strategy to our PO methods. 

%% file: sections/03-simulation.tex
\section{Simulation study} \label{sec-sim}

In this section, we assess the performance of the proposed PO methods in comparison to already existing approaches by means of a simulation study. 
For this, we follow the ADEMP (aims, data-generating mechanisms, estimands and other targets, methods, performance measures) methodology by \citet{morris2019a}, which is in spirit closely related to the clinical scenario evaluation framework by \citet{benda2010}. 
After presenting the design of the study, we also elucidate computational details regarding the different methods and their implementations. 
Eventually, we discuss the results of the study.

\subsection{Design} \label{sec-sim-design} 

\textit{Aims}\\
The aim of this simulation study is to assess the performance of the proposed PO methods for estimating and testing the RMST difference between two samples in a setup as it has been described in \secref{sec-methods}. 
Here, the focus is on scenarios with small or moderate sample sizes as it is known that some of the existing approaches suffer from an inflated type I error in these settings. 

\noindent
\textit{Data-generating mechanisms}\\
The data-generating mechanisms are, in principle, adopted from the simulation study by \citet{ditzhaus2023}, including five factors being varied: 
The event time distributions, the censoring time distributions, the base allocation of the sample size, the sample multiplier as well as the true effect size, i.e. the RMST difference. 
For the effect size $\Delta = \mu_1(t^*) - \mu_0(t^*)$  we considered  $\Delta \in \{0, 1.5\}$, reflecting scenarios under the null and under the alternative hypothesis, respectively. 
In addition, we investigate the following three pairs of event time distributions, which are named and parametrized as in \cite{ditzhaus2023}: 
\begin{itemize}
  \item[S1] Exponential distributions and proportional hazard alternatives
  \item[S7] Exponential vs. piecewise Exponential
  \item[S8] Weibull distributions with crossing curves and shape alternatives
\end{itemize}
The censoring distributions are also adopted from the same paper:
\begin{itemize}
  \item[C1] unequally Weibull distributed censoring (Weib, uneq)
  \item[C2] equally uniformly distributed censoring (Unif, eq)
  \item[C3] equally Weibull distributed censoring (Weib, eq)
\end{itemize}
Moreover, we vary the sample sizes as well as their allocation to the two treatment arms. 
Hence, the base allocations $(12, 18)$, $(15, 15)$ and $(18, 12)$ are considered for $(n_0, n_1)$. 
These base allocations are multiplied with the positive integer $K \in \{1, 2, 4, 6\}$. 
We examine $K = 6$ as we expect asymptotic arguments to take effect for this sample size.

\noindent
\textit{Estimands and other targets} \\
Although all of the methods involve the point estimation of a particular parameter, namely the two-sample RMST difference $\Delta = \muhat_1(t^*) - \muhat_0(t^*)$, they are all based on the Kaplan-Meier estimator of the survival function. 
Because of this, the point estimates of the different methods will be either exactly or nearly identical. 
The main distinction is how the statistical test of the null hypothesis in \eqref{eq:test-problem} is carried out. 
Therefore, the pair of hypotheses \eqref{eq:test-problem} is the main target of the simulation study. 

\noindent
\textit{Methods} \\
The two proposed methods based on pseudo-observations discussed in \secref{sec-methods} are of primary interest in this simulation study. 
As comparators, we include the studentized permutation test by \citet{ditzhaus2023} and the standard asymptotic test for the RMST difference (see e.g. \cite{hasegawa2020}). 
The standard asymptotic test is frequently applied in practice for RMST-based inference and therefore has the role of a baseline method that must be improved. 
On the other hand, we consider the studentized permutation test as the current gold standard for this problem and view it as a serious contender. 
Other approaches that we could have considered include the unstudentized permutation test by \citet{horiguchi2020a} and the approach based on empirical likelihood ratios by \citet{zhou2021}. 
However, based on the simulation results by \citet{ditzhaus2023}, we think that they do not fit in any of the two mentioned categories and therefore do not assess them in our study. 

\noindent
\textit{Performance measures} \\
Given the null hypothesis in \eqref{eq:test-problem} as the target of the simulation study and the original problem that the standard asymptotic test has an inflated type I error rate, the empirical type I error is also the primary performance measure. 
Corresponding secondary performance measures are the power of the respective test as well as the coverage of the associated confidence interval. \\

\begin{figure}[H]
\centering
\includegraphics[width=0.9\textwidth]{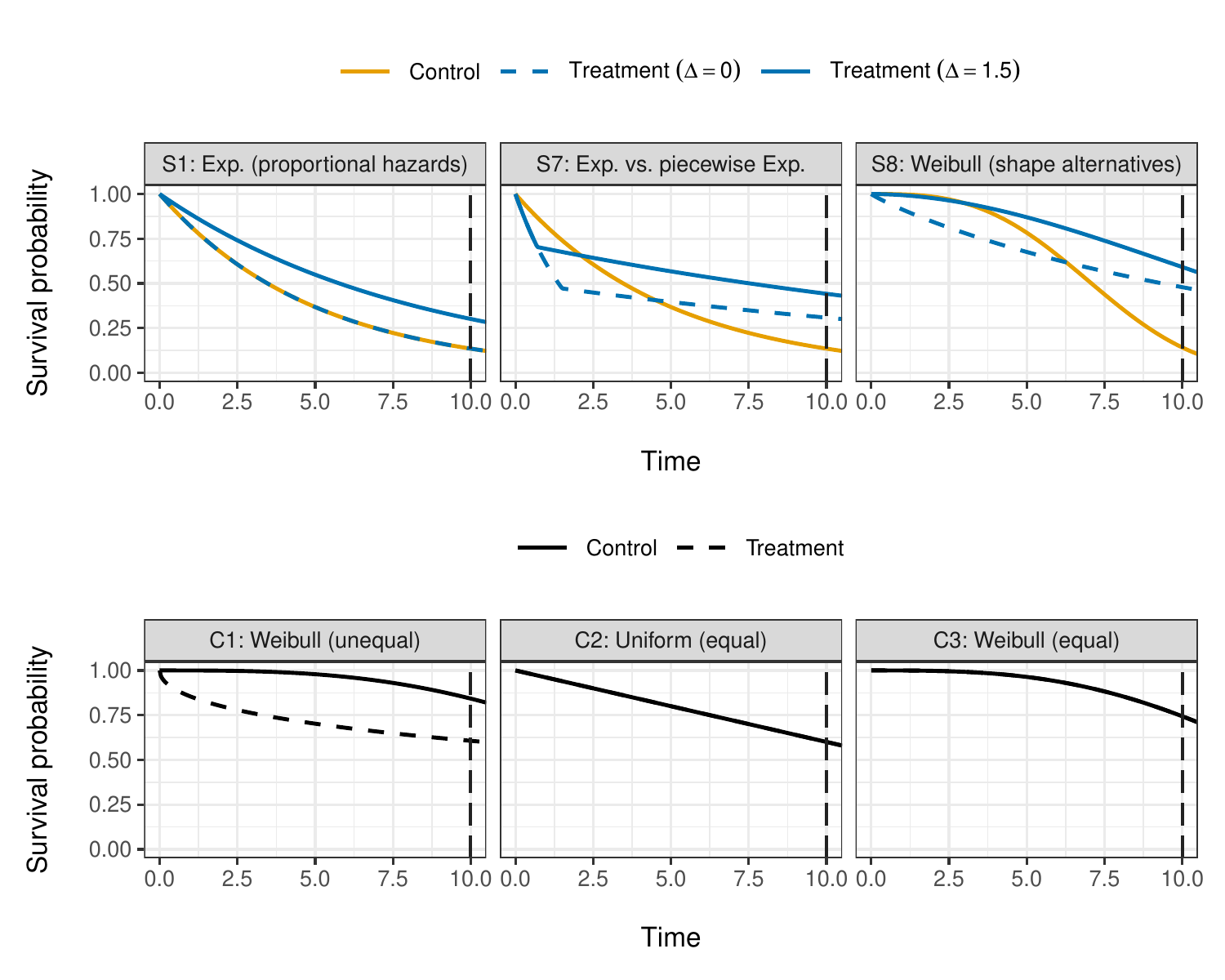}
\caption{Simulation models for the event (top) and censoring (bottom) times. The vertical dashed line indicates the cutoff time point at $t^* = 10$ for which the RMST differences $\Delta = 0$ and $\Delta = 1.5$ hold, respectively.}
\label{fig:sim-models}
\end{figure}

A summary of all design factors and an illustration of the survival and censoring models are given in \tabref{tab:tbl-simdesign} and \figref{fig:sim-models}, respectively.

\input{tables/tbl01-simdesign}

\subsection{Computational details} 

For each of the scenarios described in Section \ref{sec-sim-design}, we generated $N_{\text{sim}} = 5000$ data. 
We regenerated the data set, whenever the RMST was inestimable for at least one of the two groups \cite{horiguchi2020a,ditzhaus2023}. 
For the resampling-based approaches, we carried out $B = 2000$ resampling iterations. 
The nominal significance level $\alpha$ was set to $5\%$.

All computations were carried out using the \texttt{R} programming language in version 4.3.0 \cite{R-base} on the high-performance computing cluster of the GWDG in Göttingen.\footnote{\url{https://gwdg.de/hpc/systems/scc/}} 
The asymptotic test as well as the studentized permutation test were implemented by ourselves based on code supplied by Marc Ditzhaus. 
For the PO approaches, we used the \textit{rmeanglm()} function from the \textit{eventglm} package \cite{sachs2022}. 
Using this function, we implemented custom "modules" for calculating the pseudo-observations before estimating the regression model. 
In particular, for the IJ pseudo-observations we used the \textit{pseudo()} function from the \textit{survival} package in \textit{R} \cite{R-survival}. 
Further direct and indirect package dependencies were required for running the simulation study, all of which were tracked and managed using the \textit{renv} package \cite{R-renv}, contributing to the reproducibility of the results. 
The code for the simulation study is publicly available at GitHub.\footnote{\url{https://github.com/DavidJesse21/rmst-small-samples}}

\subsection{Results}

\input{tables/tbl02-type_I_err}

The main results concerning the type I error rate are presented in \tabref{tab:tbl-type1err}. 
Hence, we highlight the empirical type I errors that lie within the $95\%$ binomial confidence interval $[4.4\%; 5.6\%]$ around the nominal level $\alpha = 5\%$, derived from the $N_{\text{sim}} = 5000$ simulation repetitions, and consider these occasions as successes. 
First of all, we can confirm the findings by \citet{ditzhaus2023} and \citet{horiguchi2020a} regarding the standard asymptotic test as well as the studentized permutation test. 
Thus, the former makes too liberal test decisions in the vast majority of scenarios while the latter generally succeeds in addressing this issue. 
These observations are particularly pronounced for scenarios with very small sample sizes ($K = 1$). 
Moving to our two proposed PO methods, we can conclude that they provide valid alternatives for RMST-based inference that (a) should be preferred over the standard asymptotic test and (b) are competitive to the studentized permutation method. 
In total, our tests could control the type I error in 88 (asymptotic) and 90 (bootstrap) scenarios. 
The studentized permutation test is only slightly superior in settings with very small sample sizes ($K = 1$). 
On the one hand, the asymptotic PO approach can suffer from a little too liberal behavior in these scenarios, while, on the other hand, the bootstrap approach can be a bit conservative at the same time. 
For sample sizes larger than that, however, our proposed methods exhibit an excellent performance on a similar level as the studentized permutation test. 

\begin{figure}[hbtp]
\centering
\includegraphics[width=0.9\textwidth]{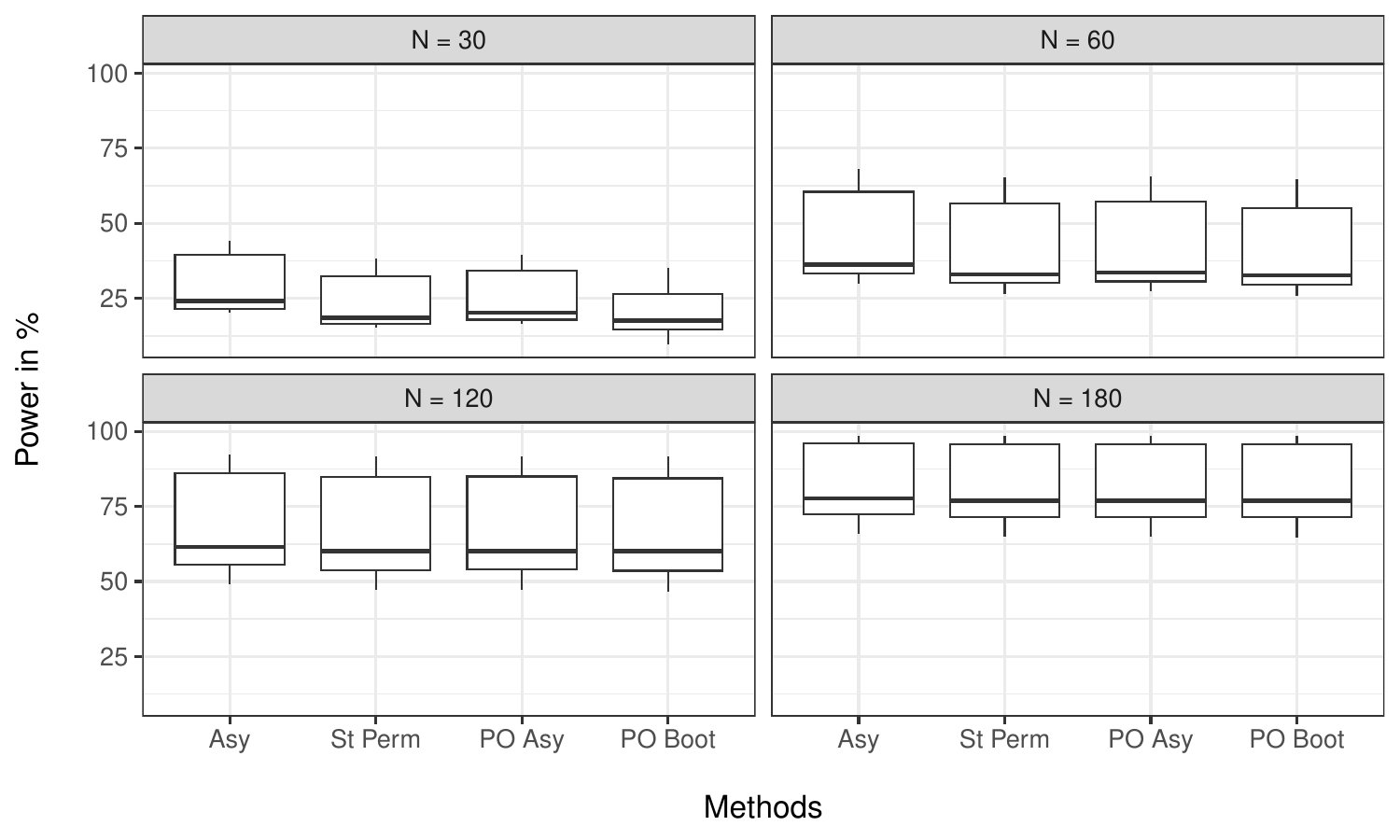}
\caption{Power values of different methods in \% (nominal level $\alpha = 5\%$) aggregated by the total sample sizes ($N = n_0 + n_1$).}
\label{fig:power}
\end{figure}

In \figref{fig:power}, we provide a high level presentation of the simulation results concerning the power of the different tests. 
Here, we discuss these results only briefly, as the observed differences between the methods are mostly minimal and therefore of minor practical relevance. 
Nonetheless, detailed numeric results are also provided in supplement S1. 
Overall, the standard asymptotic test has the largest power of all methods. 
The greatest differences can be observed for a total of 30 samples.  
However, this comes at the cost of an inflated type I error. 
Therefore, it is hard to assess the higher power of the asymptotic test as an advantage. 
Moreover, we can depict a small tendency of the bootstrap method of having a smaller power than the other approaches. 
This is in line with its conservative behavior that we have already noted in the discussion of the type I error control. 
Yet, for a total of 60 observations, these differences already begin to diminish. 

\begin{figure}[hbtp]
\centering
\includegraphics[width=0.9\textwidth]{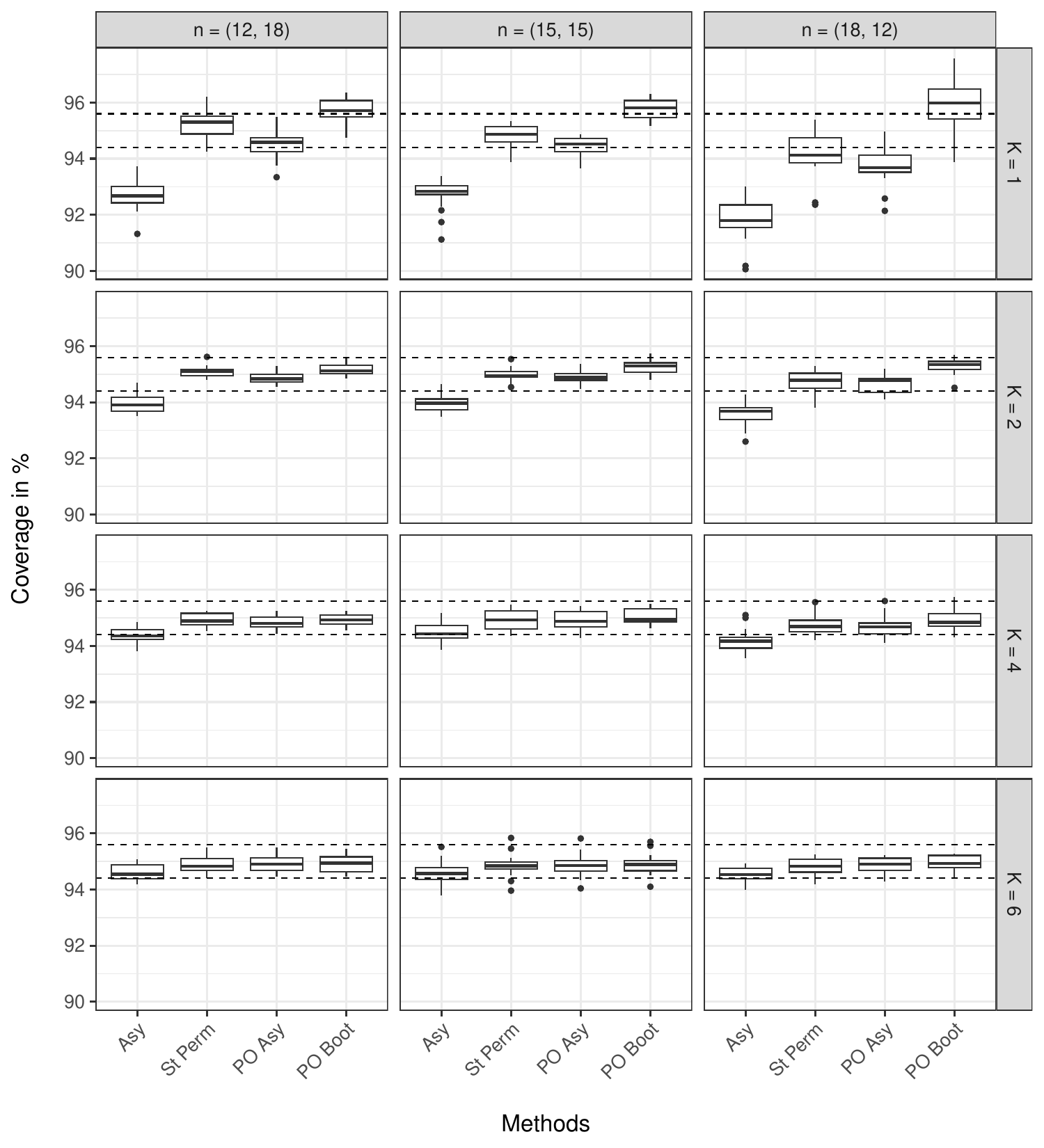}
\caption{Confidence interval coverage of different methods in \% (nominal level $\alpha = 5\%$) aggregated by sample allocations ($(n_0, n_1)$) and their multipliers ($K$). The dashed lines depict the 95\% binomial confidence interval $[94.4\%, 95.6\%]$.}
\label{fig:coverage}
\end{figure}

Finally, we discuss the empirical coverage rates of the confidence intervals associated with each method. 
These results include both, null and alternative scenarios, and are presented in \figref{fig:coverage}. 
There, we distinguish not only between the total sample sizes (rows) but also between their allocations (columns). 
Similar to the results concerning the type I error rate, we depict the 95\% binomial confidence interval around the nominal coverage rate (95\%) based on the 5000 simulation samples obtained. 
In general, the results for the confidence intervals underline our previous findings for the type I error rate when it comes to comparing the different methods with each other. 
Consequently, the standard asymptotic method fails to provide a confidence interval that maintains the nominal coverage in most scenarios. 
Overall, the studentized permutation confidence interval shows the best performance, which can again be attributed to scenarios with very small ($K = 1$) sample sizes. 
Likewise, the PO methods provide competitive confidence intervals that are able to maintain the nominal coverage in the majority of scenarios. 

%% file: tables/tbl01-simdesign.tex
\begin{table}[H]

\caption{\label{tab:tbl-simdesign}Factors and their levels for the data-generating mechanisms used in the simulation study.}
\centering
\small
\begin{tabular}[t]{ll}
\headrow
Factor & Level\\
\hiderowcolors
 & S1: Exponential distributions\\
Survival models & S7: Exponential and piecewise exponential distributions with crossing curves\\
 & S8: Weibull distributions with crossing curves and shape alternatives\\
\hline
 & C1: unequal Weibull\\
Censoring models & C2: equal uniform\\
 & C3: equal Weibull\\
\hline
RMST difference ($\Delta$) & 0; 1.5\\
\hline
Sample allocations & (12, 18); (15, 15); (18, 12)\\
\hline
Sample multipliers ($K$) & 1; 2; 4; 6\\
\hline
\end{tabular}
\end{table}

%% file: tables/tbl02-type_I_err.tex
\begin{table}

\caption{\label{tab:tbl-type1err}Type I error rates of different methods in \% (nominal level $\alpha = 5\%$). The values inside the binomial confidence interval $[4.4\%, 5.6\%]$ are printed bold.}
\centering
\begin{threeparttable}
\footnotesize
\begin{tabular}[t]{l>{}c|ccc>{}c|ccc>{}c|cccc}
\headrow
\multicolumn{1}{c}{\textbf{ }} & \multicolumn{1}{c}{\textbf{ }} & \multicolumn{4}{c}{\textbf{$N = K \cdot (12, 18)$}} & \multicolumn{4}{c}{\textbf{$N = K \cdot (15, 15)$}} & \multicolumn{4}{c}{\textbf{$N = K \cdot (18, 12)$}} \\
\headrow
Censoring & K & Asy & Perm & PO1 & PO2 & Asy & Perm & PO1 & PO2 & Asy & Perm & PO1 & PO2\\
\hiderowcolors
\multicolumn{14}{l}{S1: Exponential distributions}\\
\hline
\hspace{1em}un. W. & 1 & 7.0 & \textbf{4.6} & \textbf{5.2} & \textbf{4.4} & 7.2 & \textbf{5.4} & 5.8 & \textbf{4.6} & 8.3 & 5.9 & 6.4 & \textbf{4.6}\\
\hspace{1em} & 2 & 6.3 & \textbf{4.8} & \textbf{5.2} & \textbf{4.9} & 5.9 & \textbf{5.1} & \textbf{5.1} & \textbf{4.8} & 6.2 & \textbf{5.1} & \textbf{5.2} & \textbf{4.6}\\
\hspace{1em} & 4 & \textbf{5.3} & \textbf{4.8} & \textbf{4.9} & \textbf{4.8} & \textbf{5.0} & \textbf{4.6} & \textbf{4.6} & \textbf{4.7} & 5.7 & \textbf{5.2} & \textbf{5.2} & \textbf{5.0}\\
\hspace{1em} & 6 & \textbf{5.0} & \textbf{4.5} & \textbf{4.5} & \textbf{4.5} & \textbf{5.2} & \textbf{5.0} & \textbf{5.0} & \textbf{5.0} & \textbf{ 5.1} & \textbf{4.9} & \textbf{4.8} & \textbf{4.8}\\
\hline
\hspace{1em}eq. U. & 1 & 8.7 & 5.8 & 6.7 & \textbf{5.1} & 7.3 & \textbf{4.8} & \textbf{5.5} & \textbf{4.4} & 8.5 & \textbf{5.5} & 6.4 & \textbf{4.9}\\
\hspace{1em} & 2 & 6.4 & \textbf{5.2} & \textbf{5.4} & \textbf{4.9} & 5.9 & \textbf{4.9} & \textbf{4.9} & \textbf{4.7} & 6.4 & \textbf{5.0} & \textbf{5.2} & \textbf{4.8}\\
\hspace{1em} & 4 & 6.1 & \textbf{5.4} & \textbf{5.4} & \textbf{5.3} & \textbf{5.4} & \textbf{4.8} & \textbf{4.8} & \textbf{4.9} & 5.8 & \textbf{5.4} & \textbf{5.3} & \textbf{5.2}\\
\hspace{1em} & 6 & \textbf{5.1} & \textbf{5.1} & \textbf{4.9} & \textbf{5.0} & 5.7 & \textbf{5.4} & \textbf{5.4} & \textbf{5.4} & \textbf{ 5.2} & \textbf{4.9} & \textbf{4.9} & \textbf{4.9}\\
\hline
\hspace{1em}eq. W. & 1 & 7.9 & \textbf{5.5} & 6.1 & \textbf{5.3} & 7.2 & \textbf{5.0} & \textbf{5.4} & \textbf{4.7} & 7.7 & \textbf{5.3} & 6.1 & \textbf{4.9}\\
\hspace{1em} & 2 & 5.8 & \textbf{4.9} & \textbf{5.0} & \textbf{5.0} & 5.9 & \textbf{5.1} & \textbf{5.2} & \textbf{5.0} & 5.7 & \textbf{4.7} & \textbf{4.8} & \textbf{4.5}\\
\hspace{1em} & 4 & 5.8 & \textbf{5.3} & \textbf{5.3} & \textbf{5.3} & \textbf{4.8} & \textbf{4.5} & \textbf{4.6} & \textbf{4.5} & 5.7 & \textbf{5.0} & \textbf{5.2} & \textbf{5.1}\\
\hspace{1em} & 6 & 5.7 & \textbf{5.5} & \textbf{5.5} & \textbf{5.5} & \textbf{5.4} & \textbf{5.3} & \textbf{5.2} & \textbf{5.2} & \textbf{ 5.1} & \textbf{4.8} & \textbf{4.8} & \textbf{4.8}\\
\hline
\multicolumn{14}{l}{S7: Exponential and piecewise exponential distributions with crossing curves}\\
\hline
\hspace{1em}un. W. & 1 & 6.5 & 3.8 & \textbf{4.6} & 3.7 & 6.9 & \textbf{5.0} & \textbf{5.3} & 3.9 & 8.3 & 6.2 & 6.3 & 4.0\\
\hspace{1em} & 2 & 6.2 & \textbf{4.9} & \textbf{5.3} & \textbf{4.9} & 6.3 & \textbf{5.1} & \textbf{5.1} & \textbf{4.6} & 7.1 & 6.2 & 5.9 & \textbf{5.0}\\
\hspace{1em} & 4 & \textbf{5.5} & \textbf{5.1} & \textbf{5.1} & \textbf{4.9} & 5.7 & \textbf{5.2} & \textbf{5.2} & \textbf{5.1} & 6.2 & \textbf{5.6} & 5.7 & \textbf{5.3}\\
\hspace{1em} & 6 & \textbf{5.1} & \textbf{4.8} & \textbf{4.9} & \textbf{4.7} & \textbf{5.5} & \textbf{5.1} & \textbf{5.2} & \textbf{5.1} & \textbf{ 5.4} & \textbf{5.1} & \textbf{5.0} & \textbf{4.7}\\
\hline
\hspace{1em}eq. U. & 1 & 6.9 & 4.1 & \textbf{5.0} & 3.9 & 7.2 & \textbf{5.2} & 5.7 & \textbf{4.6} & 7.2 & \textbf{4.8} & \textbf{5.4} & 3.3\\
\hspace{1em} & 2 & 6.5 & \textbf{5.1} & \textbf{5.4} & \textbf{5.1} & 6.5 & \textbf{5.3} & \textbf{5.4} & \textbf{5.0} & 6.2 & \textbf{4.9} & \textbf{5.1} & \textbf{4.6}\\
\hspace{1em} & 4 & 5.8 & \textbf{5.1} & \textbf{5.3} & \textbf{5.1} & 5.7 & \textbf{5.1} & \textbf{5.1} & \textbf{5.1} & 5.8 & \textbf{5.2} & \textbf{5.3} & \textbf{5.0}\\
\hspace{1em} & 6 & \textbf{5.3} & \textbf{4.9} & \textbf{4.8} & \textbf{4.8} & \textbf{4.5} & 4.2 & 4.2 & 4.3 & \textbf{ 5.6} & \textbf{5.3} & \textbf{5.3} & \textbf{5.2}\\
\hline
\hspace{1em}eq. W. & 1 & 7.4 & \textbf{5.4} & 6.1 & \textbf{5.2} & 6.7 & \textbf{4.9} & \textbf{5.3} & \textbf{4.4} & 8.2 & 6.1 & 6.5 & 3.8\\
\hspace{1em} & 2 & 5.8 & \textbf{4.7} & \textbf{5.0} & \textbf{4.8} & \textbf{5.5} & \textbf{4.7} & \textbf{4.8} & \textbf{4.5} & 6.3 & \textbf{5.3} & \textbf{5.3} & \textbf{4.8}\\
\hspace{1em} & 4 & \textbf{5.1} & \textbf{4.8} & \textbf{4.7} & \textbf{4.8} & \textbf{5.5} & \textbf{5.1} & \textbf{5.1} & \textbf{5.0} & 5.9 & \textbf{5.4} & \textbf{5.5} & \textbf{5.4}\\
\hspace{1em} & 6 & \textbf{5.1} & \textbf{4.9} & \textbf{4.9} & \textbf{4.9} & \textbf{5.6} & \textbf{5.2} & \textbf{5.4} & \textbf{5.5} & \textbf{ 5.5} & \textbf{5.3} & \textbf{5.3} & \textbf{5.3}\\
\hline
\multicolumn{14}{l}{S8: Weibull distributions with crossing curves and shape alternatives}\\
\hline
\hspace{1em}un. W. & 1 & 7.0 & \textbf{4.6} & \textbf{5.2} & 3.9 & 8.9 & 6.1 & 6.2 & 4.0 & 9.8 & 7.6 & 7.4 & 3.6\\
\hspace{1em} & 2 & \textbf{5.3} & \textbf{4.4} & \textbf{4.7} & \textbf{4.4} & 6.4 & \textbf{5.4} & \textbf{5.3} & \textbf{4.7} & 7.4 & 6.1 & 5.9 & \textbf{4.7}\\
\hspace{1em} & 4 & \textbf{5.5} & \textbf{4.8} & \textbf{4.9} & \textbf{4.9} & 5.7 & \textbf{5.5} & \textbf{5.4} & \textbf{5.1} & 6.0 & \textbf{5.5} & \textbf{5.5} & \textbf{4.9}\\
\hspace{1em} & 6 & \textbf{5.1} & \textbf{4.8} & \textbf{4.7} & \textbf{4.6} & \textbf{4.8} & \textbf{4.5} & \textbf{4.6} & \textbf{4.4} & 5.9 & \textbf{5.6} & \textbf{5.5} & \textbf{5.3}\\
\hline
\hspace{1em}eq. U. & 1 & 7.4 & \textbf{4.6} & \textbf{5.3} & 4.1 & 7.3 & \textbf{5.1} & \textbf{5.3} & 3.7 & 8.8 & 6.3 & 6.7 & 3.3\\
\hspace{1em} & 2 & 6.3 & \textbf{5.1} & \textbf{5.3} & \textbf{4.9} & \textbf{5.6} & \textbf{4.7} & \textbf{4.7} & 4.3 & 6.6 & \textbf{5.5} & 5.7 & \textbf{4.5}\\
\hspace{1em} & 4 & 5.7 & \textbf{5.1} & \textbf{5.1} & \textbf{5.1} & \textbf{5.1} & \textbf{4.7} & \textbf{4.6} & \textbf{4.6} & 5.8 & \textbf{5.1} & \textbf{5.2} & \textbf{4.9}\\
\hspace{1em} & 6 & \textbf{5.5} & \textbf{5.2} & \textbf{5.2} & \textbf{5.2} & \textbf{5.6} & \textbf{5.2} & \textbf{5.2} & \textbf{5.0} & \textbf{ 5.4} & \textbf{5.1} & \textbf{5.0} & \textbf{4.8}\\
\hline
\hspace{1em}eq. W. & 1 & 6.3 & 4.1 & \textbf{4.5} & 3.9 & 7.2 & \textbf{5.4} & \textbf{5.5} & 4.0 & 8.1 & 6.1 & 6.3 & 4.0\\
\hspace{1em} & 2 & 6.3 & \textbf{5.1} & \textbf{5.4} & \textbf{5.1} & 6.0 & \textbf{5.0} & \textbf{5.0} & \textbf{4.8} & 6.3 & \textbf{5.5} & \textbf{5.6} & \textbf{4.8}\\
\hspace{1em} & 4 & \textbf{5.4} & \textbf{4.8} & \textbf{4.9} & \textbf{4.9} & 5.9 & \textbf{5.4} & \textbf{5.5} & \textbf{5.2} & 5.8 & \textbf{5.3} & \textbf{5.3} & \textbf{5.2}\\
\hspace{1em} & 6 & 5.8 & \textbf{5.6} & \textbf{5.5} & \textbf{5.5} & 6.2 & 6.0 & 6.0 & 5.9 & \textbf{ 5.6} & \textbf{5.4} & \textbf{5.3} & \textbf{5.1}\\
\hline
\end{tabular}
\begin{tablenotes}
\item \textit{Abbreviations:} Asy, asymptotic test; Perm, studentized permutation test; PO1, pseudo-observations asymptotic; PO2, pseudo-observations bootstrap; un. W., unequal Weibull censoring; eq. U., equal uniform censoring; eq. W., equal Weibull censoring.
\end{tablenotes}
\end{threeparttable}
\end{table}

%% file: sections/04-example.tex
\section{Empirical example} \label{sec-example}

It is considered good practice to complement statistical simulation studies with applications to real-world data sets, e.g. to illustrate if or to what extent the choice of method matters in practical analyses \cite{friedrich2024}. 
Therefore, we want to apply the methods we have compared in the simulation study to a data set from the medical literature to see how the obtained results differ. 
In addition, we want to look at an aspect of our proposed PO methods that we have not systematically covered in the simulation study, that is the possibility to conduct covariate-adjusted analyses. 
This facet is particularly interesting as it is known from other contexts that the incorporation of prognostic covariates into the analysis can increase the precision of treatment effect estimates and consequently increase the power of associated tests \cite{kahan2014}. 

\begin{figure}[hbtp]
\centering
\includegraphics[width=0.8\textwidth]{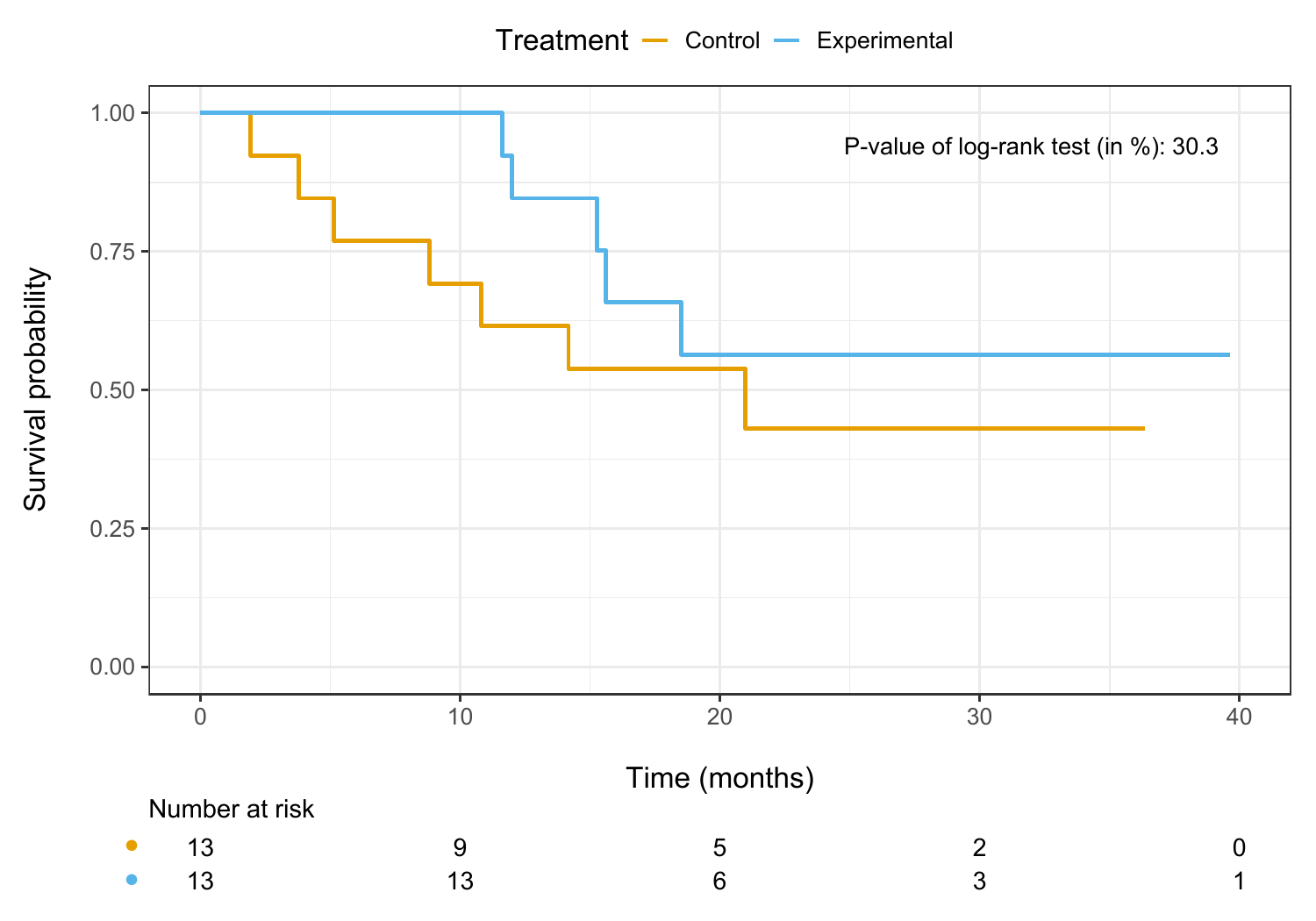}
\caption{Kaplan-Meier estimates of the groupwise survival functions for the data from \citet{edmonson1979}. The p-value of the log-rank test is displayed in the top-right corner of the plot.}
\label{fig:edmonson}
\end{figure}

The data set we look at is from a study by \citet{edmonson1979}, in which patients suffering from ovarian cancer were randomized to either one of two treatments. 
The first treatment regimen consisted of cyclophosphamide only, whereas the other group received adriamycin in addition to cyclophosphamide. 
Additionally, the age of the patients as well as their ECOG (Eastern Cooperative Oncology Group) performance score at baseline are recorded. 
The data set is publicly available in the R  \texttt{survival} package \cite{R-survival} as the "ovarian" data set. 
\figref{fig:edmonson} shows the Kaplan-Meier estimate of the survival function for each of the two treatment arms. 
Visually, a positive effect favoring the experimental treatment might be expected as the estimated survival probability is higher than that of the control arm throughout the entire study period. 
Nonetheless, the difference in survival probabilities decreases towards the end of the study, making this deduction less clear. 
It must also be considered that the sample size is relatively small with 13 patients in each treatment arm only. 
The log-rank test outputs a p-value of 30.3\%, thus retaining (i.e. not rejecting) the null hypothesis that the survival functions of both groups are the same and that there is no treatment effect. 

Although a violation of the PH assumption is not obvious, we now analyze the data using the RMST-based methods we have contrasted in \secref{sec-sim}. 
We also augment this list of methods with covariate-adjusted versions of the PO approaches. 
For these, we use the two aforementioned variables, the age of the patients and their ECOG performance score. 
For the cutoff time points $t^*$, we use three different values in order to highlight the dependence of the conclusions on this choice. 
Precisely, we look at $t^* \in \{15,\, 20,\, 25\}$ months. 
We set the nominal significance level $\alpha$ to $5\%$ for a two-sided test and use $B = 5000$ resampling iterations for the permutation and bootstrap approaches. 
We want to emphasize that no adjustments for multiple testing, e.g. regarding the different cutoff time points are made. 

\begin{figure}[hbtp]
\centering
\includegraphics[width=0.9\textwidth]{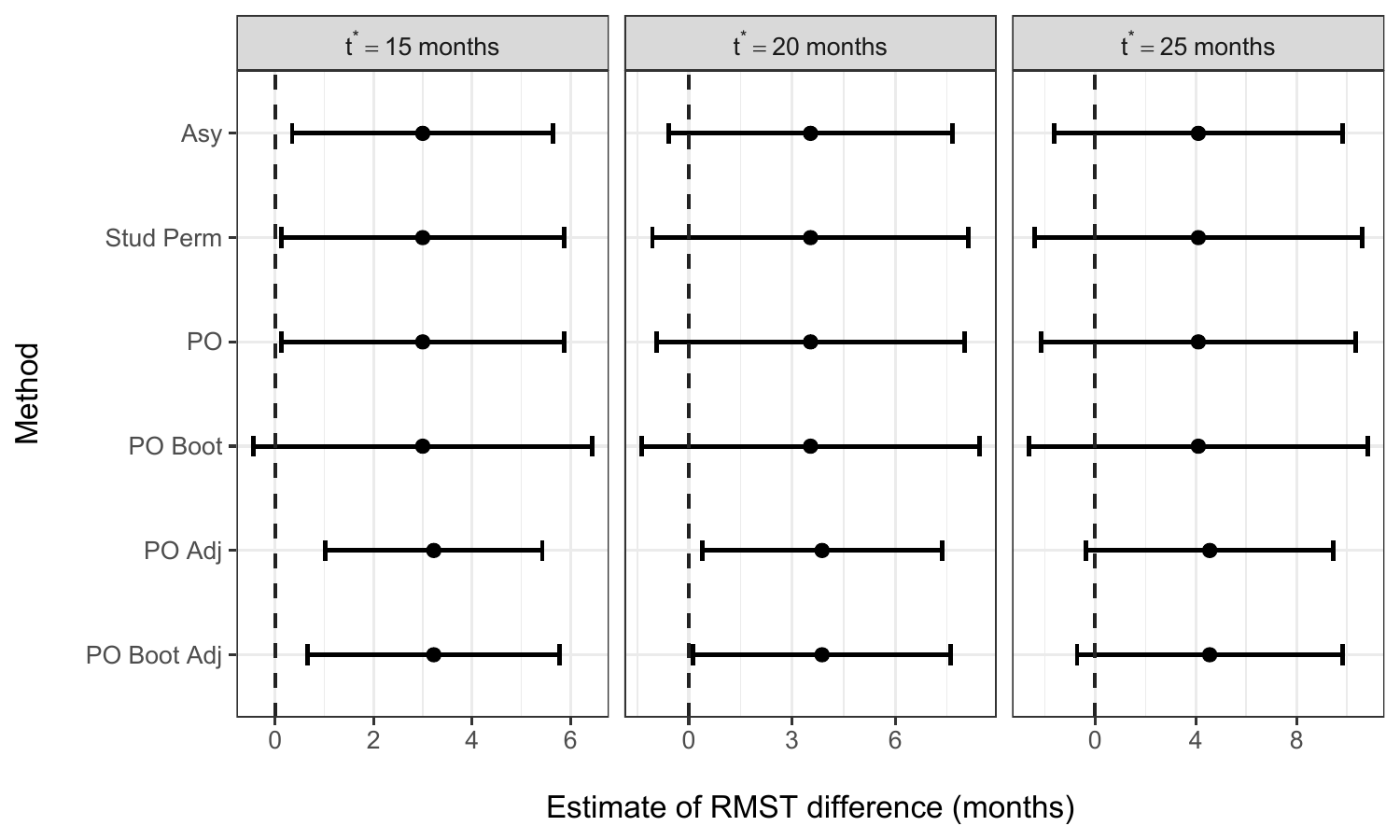}
\caption{Point estimates of the RMST difference and 95\% confidence intervals for the data from \citet{edmonson1979}. The dashed lines highlight an RMST difference of 0, i.e. no treatment difference.}
\label{fig:edmonson-res}
\end{figure}

\figref{fig:edmonson-res} shows the results in terms of the point estimates and the corresponding $95\%$ confidence intervals. 
For $t^* = 15$ months, we find that all unadjusted approaches except for the bootstrap method estimate a treatment effect that is statistically significant at the level of $5\%$. 
While the confidence intervals of the studentized permutation and the asymptotic PO methods are quite similar, we can see that the ones of the standard asymptotic and the PO bootstrap approach are noticeably narrower and wider, respectively. 
Things change, however, when we consider the adjusted analyses. 
On the one hand, the point estimates of the treatment effect become somewhat larger and, on the other hand, the confidence intervals shrink noticeably. 
The described patterns and differences regarding the methods remain the same for the other two cutoff time points. 
However, the unadjusted approaches now all yield the conclusion that the estimated effect is not statistically significant, while for $t^* = 20$ months the effect is still deemed significant by the adjusted methods. 
For $t^* = 25$ months, however, none of the methods give evidence for an RMST difference that is larger than $0$.

%% file: sections/05-conclusion.tex
\section{Conclusion} \label{sec-discuss}

In this paper, we contributed to the ongoing discussion about handling small sample sizes for RMST-based inference in two-sample comparisons \cite{horiguchi2020a,ditzhaus2023}. 
To this end, we proposed two methods based on PO regression models and compared them to the standard asymptotic test \cite{hasegawa2020} and the studentized permutation test \cite{ditzhaus2023} in a simulation study. 
We can conclude that the PO methods perform better than the standard asymptotic test and almost as well as the studentized permutation test in terms of type I error rate control. 
Thus, we provide methods for RMST-based inference that should be preferred over the standard asymptotic approach, which is commonly applied in this context. 
An aspect that makes the PO methods even more attractive is their flexibility to adapt to other settings, which they inherit from the class of GLMs. 
For instance, \citeauthor{ditzhaus2023} mention that their studentized permutation method can also be applied to situations with competing risks or related summary measures such as window mean survival time \cite{paukner2021}. 
This is also easily possible with the PO approaches as the only aspect of the method that needs to be changed is the marginal estimator used for calculating the pseudo-observations \cite{sachs2022}. 
Moreover, \citet{munko2024} recently introduced RMST-based tests for multiple comparisons, also employing the studentized permutation methodology. 
The PO approaches might be equally applicable to this type of problem. 
The biggest advantage of the PO methods, however, can be considered the possibility to perform covariate-adjusted analyses, especially with continuous covariables. 
When prognostic variables are adjusted for in the analysis, this can minimize the risk of bias and lead to a gain in the precision of treatment effect estimates and to a higher power of the associated tests \cite{kahan2014,karrison2018}. 
Although we have touched on this aspect in \secref{sec-example}, this idea needs to be investigated more elaborately in future research.